\documentclass[twoside,a4paper,11pt]{sea10}
\usepackage{graphicx}
\usepackage{hyperref}
\usepackage{movie15}
\usepackage{bib}
\newcommand{\HII}{H {\scshape{ii}} }
\newcommand\sun{\odot}
\begin{document}
\pagenumbering{arabic}
\pagestyle{myheadings}
\thispagestyle{empty}
{\flushleft\includegraphics[width=\textwidth,bb=58 650 590 680]{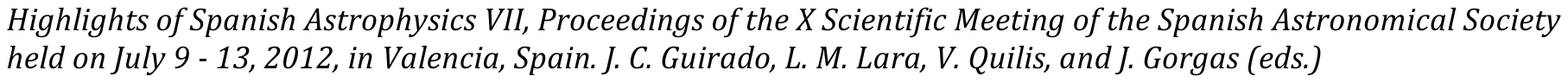}}
\vspace*{0.2cm}
\begin{flushleft}
{\bf {\LARGE
%
Environmental imprint on galaxy chemical enrichment
}\\
\vspace*{1cm}
%
Vasiliki Petropoulou$^{1}$,
J.M. V\'ilchez$^{2}$, 
and 
J. Iglesias-P\'aramo$^{2}$
%
}\\
\vspace*{0.5cm}
%
$^{1}$
INAF - Osservatorio Astronomico di Brera\\
$^{2}$
Instituto de Astrof\'isica de Andaluc\'ia  - CSIC\\

%
\end{flushleft}
%
\markboth{
Short version of the paper title
}{ 
%
V. Petropoulou et al. 
%
}
\thispagestyle{empty}
\vspace*{0.4cm}
\begin{minipage}[l]{0.09\textwidth}
\ 
\end{minipage}
\begin{minipage}[r]{0.9\textwidth}
\vspace{1cm}
\section*{Abstract}{\small
%

Recent results are presented on the metal enrichment of low-mass star-forming (SF) galaxies in local Universe clusters. The environmental effects on the chemical evolution of these galaxies are discussed. We have used spectroscopic data from the SDSSIII-DR8 and we have derived the gas-phase O/H and N/O abundances. We have then examined the Mass-Metallicity Relation (MZR) of this sample of cluster galaxies, and we have found well defined sequences. A flattening of the slope of these sequences has been observed for galaxies located in the core of the two more massive clusters of the sample, suggesting that the effect of the cluster environment depends both on the galaxy mass and the host cluster mass. Based on these results we explore cluster-specific effects (e.g. ram-pressure stripping, pressure confinement etc), predicted by hydrodynamic models, capable of yielding the observed mass-dependent enhancement of the metallicity.

%
\normalsize}
\end{minipage}
%
%
%
\section{Introduction \label{intro}}

The starting point of this investigation is that we know that galaxy evolution is connected with the environment. As galaxies enter to the cluster environment they have their star formation histories (SFH) affected, with a possible enhancement \cite{Kronberger2008,Kapferer2009b} preceeding an eventual shutting down of their star formation \cite{Haines2007}. Additionally, they have their gas component severely affected, as confirmed by observations of their atomic and ionized gas \cite{BravoAlfaro2000,Koopmann2004}. 

As galaxy metallicity is a physical parameter directly connected with both the SFH of a galaxy and the gas exchange with the environment, the question that we propose to investigate is whether we can see the effect of the environment imprinted on galaxy metallicity. 

According to previous works on the gas metallicity of SF dwarf  galaxies in local Universe clusters, the effect of the environment seems to be relevant. In Virgo dIrr, the gas metallicity has not been found to show a clear trend with the environment \cite{Vilchez1995, Lee2003, Vaduvescu2007}. Some SF dwarfs in Hydra cluster  \cite{Duc2001} have been found metal-rich for their luminosities and \cite{Vaduvescu2011}, comparing the MZR of SF dwarfs in Hydra, Fornax and Virgo, have suggested that differences in the MZR seem to exist for galaxies in such different environments. 

Going to a higher mass cluster, the Hercules (A2151) cluster, in a previous work \cite{Petr2011} we found that the low-mass galaxies with the highest interstellar medium metallicities are located at the highest local densities, showing also a shift from the general MZR. Given these previous findings we focused here our attention only to the low-mass SF galaxies in a  sample of local Universe clusters.

\section{Low-mass star-forming galaxies in four local Universe clusters}

We used the SDSS spectroscopy to study the chemical abundances of a large sample  of  SF galaxies with masses from $10^8$ to $10^{10}$ $M_\sun$, located within 3$R_{200}$ distance around the center of the nearby clusters A1656 (Coma), A1367, A779, and A634. These local Universe clusters span a wide range of masses ($12.0$, $8.1$, $0.5$, and $0.7$ $\times 10^{14}$ $M_\sun$ respectively) allowing to test the role of the cluster mass in this respect.  

We note that selecting only the low-mass galaxies we avoid the aperture-bias of the SDSS  spectroscopy \cite{Kewley2005} as low-mass galaxies have not been found to show metallicity gradients \cite{vZH2006}. We used the emission line measures provided by the SDSS-DR8, derived after correcting the spectra for the underlying stellar continuum \cite{Tremonti2004,Brinchmann2004}, an important step to derive accurate metallicities. We also used the mass estimates provided by SDSS-DR8 \cite{Kauffmann2003}. 

We corrected the line fluxes for extinction and then we computed emission line ratios that we combined into commonly used diagnostic diagrams. The spectral properties of the extragalactic \HII regions have been extensively studied the last four decades \cite{McCall1985} and well defined sequences have been found among certain line ratios that our sample of galaxies appear to follow independently of their cluster membership. We then derived the chemical abundances O/H and N/O using the empirical calibrations of \cite{PMC2009} and \cite{Pettini2004}.

\begin{figure}
\center
\includegraphics[scale=0.3]{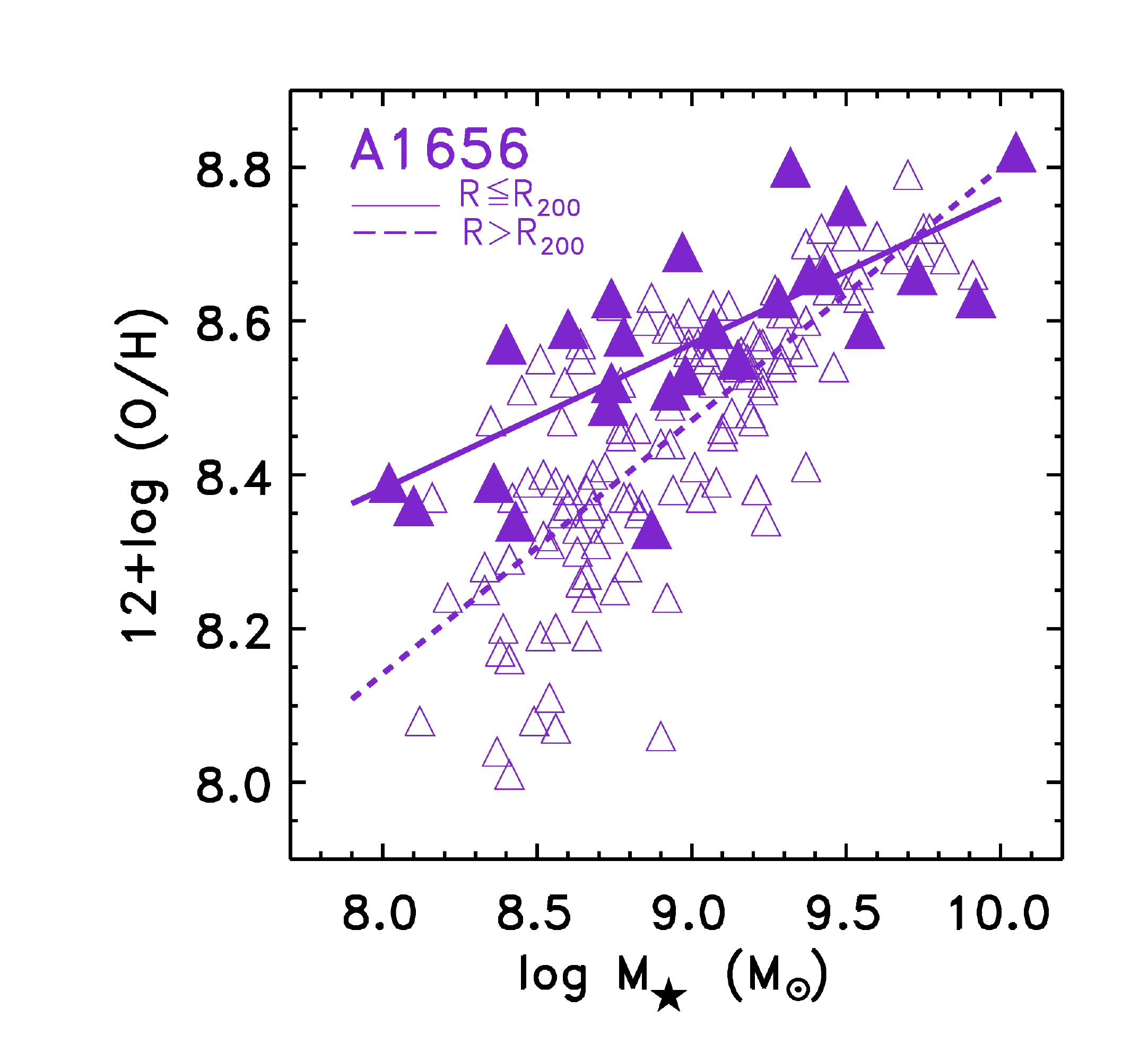}
\includegraphics[scale=0.3]{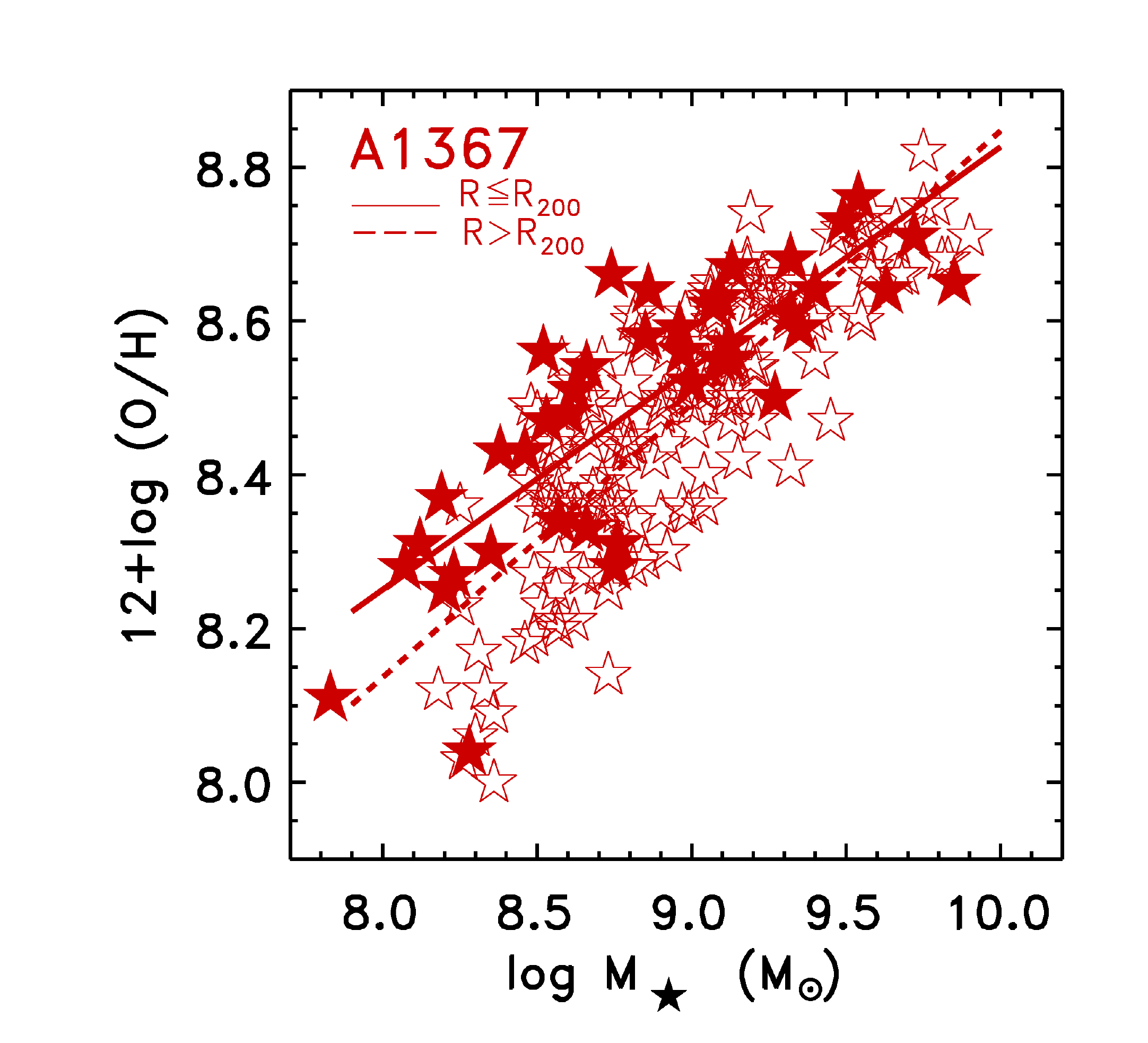}
\caption{\label{fig1} The oxygen abundance 12+log(O/H) versus stellar mass for low-mass SF galaxies in Coma (A1656, triangles), A1367 (stars). Filled symbols correspond to galaxies at $R\le R_{200}$. The continuous line is the linear fit for galaxies at $R\le R_{200}$ and the dashed line the fit for galaxies at $R>R_{200}$.}
\end{figure}

In Fig.~\ref{fig1} we plot the MZR of the low-mass SF galaxies in A1656 (left) and A1367 (right). Filled symbols correspond to galaxies located at distances  $R \le R_{200}$ while open symbols represent galaxies located at $R_{200}<R \le 3R_{200}$. We observe that at lower masses the derived metallicities cover a range of $\sim0.4$ dex while at higher masses the relation becomes tighter, appearing to form a triangle instead of a simple linear correlation. We also observe that at the same bin of mass, galaxies located inside the $R_{200}$ region of the cluster are preferentially located in the upper part of the global relation, showing higher shift for the lower masses. The same effect, though to a lower extend, was observed for A1367. In turn, we did not verify this feature for the two clusters of lower mass of our sample: A779 and A634. 

We note that we checked and rejected possible biases in the abundance derivation and in the mass estimates. We also checked that this separation is not related to the star formation rate (SFR) of the galaxies, neither the color nor the age of the galaxies (we derived these properties from SDSSIII-DR8 data). Thus we suspect that there must be a connection of this mass-dependent chemical enhancement with the cluster environment.

We perform linear fits to the MZR of the galaxies in each one of the four studied clusters and the derived value of the slope is in all cases $\sim0.3$, which is in agreement with the slope of the MZR of a sample of nearby field dwarf galaxies \cite{Lee2006} and in agreement with the predictions of  hydrodynamic models \cite{Finlator2008,Dave2011}. For A1656 and A1367 we considered also separately the cluster-core galaxies and the slopes derived show lower values ($0.19\pm0.03$ and $0.28\pm0.03$ respectively), principally for A1656, indicating a flattening of its MZR. 

This flattening of the MZR reveals that the observed chemical enhancement of the low-mass SF cluster galaxies is mass dependent: it depends on the galaxy mass, being more important for lower mass galaxies and it depends on cluster mass, being more important for the cluster of higher mass in our sample.

We also plotted the N/O versus stellar mass for all four clusters and we found nice correlations, relatively tighter than for O/H. The increased chemical abundance for A1656 low-mass SF galaxies is also observed in the N/O versus mass relation. The N/O ratio is a ``chemical clock'' \cite{Edmunds1978} indicating that if there is an environmental effect driving the observed difference in the abundances of A1656 low-mass SF galaxies, this effect should be acting at least during $10^{8}$ yr. 

\begin{figure}
\center
\includegraphics[scale=0.25]{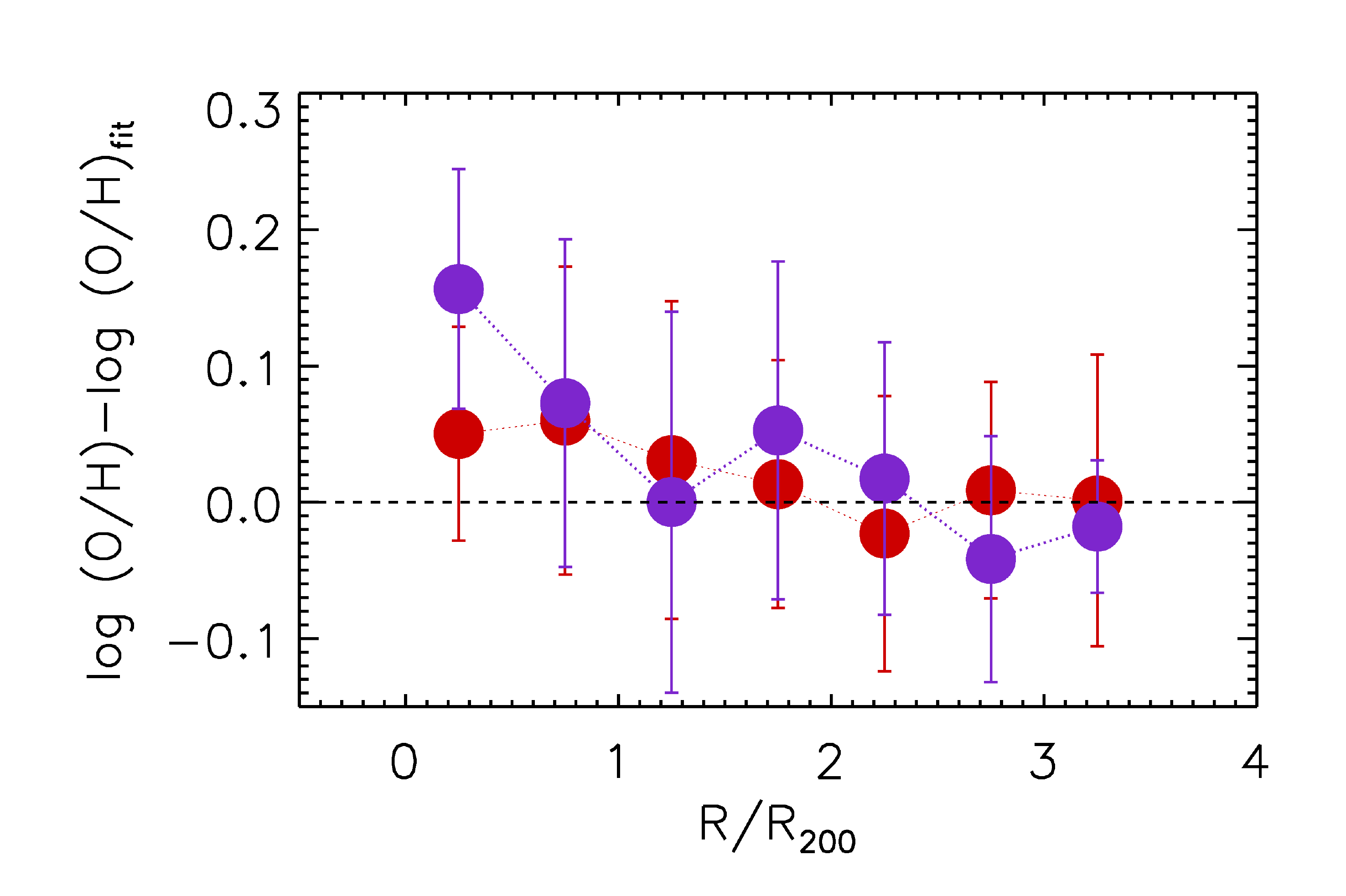}
\includegraphics[scale=0.25]{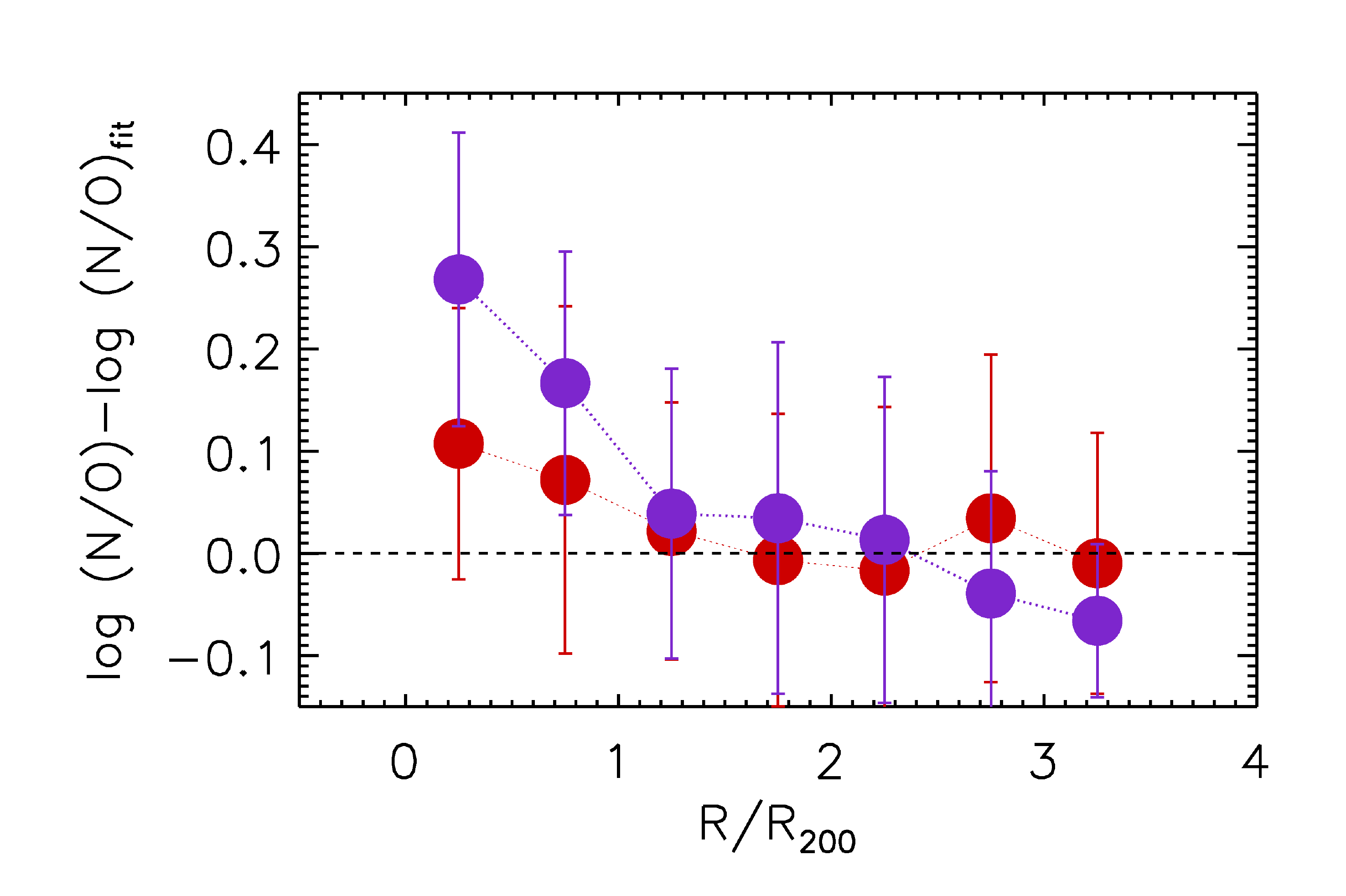}
\caption{\label{fig2} Left: The difference of the derived oxygen abundance 12+log(O/H), for each galaxy, with the oxygen abundance given by the linear fit (the dashed line in Fig.~\ref{fig1}), as a function of the cluster-centric radial distance $R/R_{200}$, in a bin of 0.5 $R_{200}$. Right: The same for log(N/O). The blue points correspond to A1565 and the red poins to A1367.}
\end{figure}

There is an important open question regarding the environmental impact on galaxy evolution, which is whether this impact is related with the cluster environment or just the high local density. We addressed this question checking, for A1656 and A1367 galaxies, the residuals in metallicity from the overall fits to the MZR, as a function of cluster-centric distance (Fig.~\ref{fig2}) and local galaxy density. We derive local galaxy density to the average of the projected distances to the fourth and fifth nearest neighbors.

We found that the metallicity enhancement is much more prominently observed as a function of cluster-centric distance, reaching a mean difference of $\sim$0.15 dex in O/H (Fig.~\ref{fig2}, left) and $\sim$0.3 dex in N/O (Fig.~\ref{fig2}, right) in the center of A1656. For A1367 lower mean differences are observed. Lower difference are observed as a function of the local galaxy density too. Thus, the relevant parameter to affect the chemical evolution should be the presence of the hot intracluster medium (ICM) gas in the central region of a cluster of the mass of A1656.

\section{The proposed scenario \label{sec}} 

The MZR was first observed as a luminosity-metallicity correlation \cite{Lequeux1979} and then it was understood to be a more fundamental MZR that has been found to hold for large samples of galaxies \cite{Tremonti2004}, spanning the whole range of stellar masses. MZR has been also observed to hold to higher redshifts, although there exist discussions on whether it is a universal relation or there is an evolution.  

Despite 30 years of efforts, the physical processes that drive this correlation are not yet fully understood. According to given models the principal mechanism are the gas outflows. Other models advocate either a systematic variation of the IMF, or the so-called ``downsizing'', or even gas inflows.

Among the most successful models are those proposed by \cite{Dave2011} and \cite{Finlator2008} that introduce an equilibrium picture and momentum driven winds. In this equilibrium picture, the gas-phase metallicity is set by a balance between inflows, outflows and star formation.

In this frame, the environment could affect any of these three parameters introduced in the equilibrium model and in any case the consequence would be the enhancement of the galaxy chemical abundance.

Inflows for example would be curtailed as a galaxy in a dense environment would loose part of its atomic gas \cite{Solanes2001} and the hot halo gas \cite{Bekki2009}  due to gas stripping mechanisms (e.g. ram-pressure stripping, starvation, tidal stripping, etc). 

The outflows would be also altered in an environment of high pressure such as in a cluster core. Hydrodynamic simulations  indicate that the pressures reached even at the outskirts of a cluster as massive as A1656 \cite{Tecce2011} would be enough to cause wind suppression \cite{Kapferer2009} and when the wind speed would reach the point to be lower than the escape velocity, the expelled material would fall back producing a wind reaccretion \cite{Oppenheimer2010}.

Finally, SFR could be temporally enhanced by the environment, due to effects such as pressure-triggered star formation \cite{Kronberger2008,Kapferer2009b}  or galaxy-galaxy interactions. 

All the previously mentioned effects are observed in a cluster and some of them are also observed in a group environment. Detailed modeling would be necessary in order to disentangle the role played by each one of these mechanisms. However, as we observed the chemical enhancement primarily for A1656 low-mass SF galaxies, we propose that the mechanisms related to the pressure of the hot ICM should be principally responsible for the observed effects.

We conclude that the enhanced metal enrichment could be produced by the combination of effects such as wind reaccretion, due to pressure confinement by the ICM, truncation of gas infall as a result of ram-pressure stripping and accelerated SFR. We close saying that the environmental impact has been an indirect way to tackle the question of the underlying physical process driving chemical evolution.

%
\small  
%
%

%

%

\begin{thebibliography}{}
\small
%

\bibitem{Bekki2009} Bekki, K.\ 2009, \mnras, 399, 2221 

\bibitem{BravoAlfaro2000} Bravo-Alfaro, H., Cayatte, V., van Gorkom, J.~H., \& Balkowski, C.\ 2000, \aj, 119, 580 



\bibitem{Brinchmann2004} Brinchmann, J., Charlot, S., White, S.~D.~M., et al.\ 2004, \mnras, 351, 1151 

\bibitem{Dave2011} Dav{\'e}, R., Finlator, K., \& Oppenheimer, B.~D.\ 2011, \mnras, 416, 1354 

\bibitem{Duc2001} Duc, P.-A., Cayatte, V., Balkowski, C., Thuan, T.~X., Papaderos, P., \& van Driel, W.\ 2001, \aap, 369, 763

\bibitem{Edmunds1978} Edmunds, M.~G., \& Pagel, B.~E.~J.\ 1978, \mnras, 185, 77P 

\bibitem{Finlator2008} Finlator, K., \& Dav{\'e}, R.\ 2008, \mnras, 385, 2181

\bibitem{Haines2007} Haines, C.~P., Gargiulo,
A., La Barbera, F., Mercurio, A., Merluzzi, P.,
\& Busarello, G.\ 2007, \mnras, 381, 7 


\bibitem{Kapferer2009} Kapferer, W., Kronberger, T., Breitschwerdt, D., et al.\ 2009, \aap, 504, 719 

\bibitem{Kapferer2009b} Kapferer, W., Sluka, C., Schindler, S., Ferrari, C., \& Ziegler, B.\ 2009, \aap, 499, 87 


\bibitem{Kauffmann2003} Kauffmann, G., 
Heckman, T.~M., White, S.~D.~M., et al.\ 2003, \mnras, 341, 33 

\bibitem{Kewley2005} Kewley, L.~J., Jansen, 
R.~A., \& Geller, M.~J.\ 2005, \pasp, 117, 227 

\bibitem{Koopmann2004} Koopmann, R.~A., \& Kenney, J.~D.~P.\ 2004, \apj, 613, 866

\bibitem{Kronberger2008} Kronberger, T., Kapferer, W., Ferrari, C., Unterguggenberger, S., \& Schindler, S.\ 2008, \aap, 481, 337 


\bibitem{Lee2003} Lee, H., McCall, M.~L., \& Richer, M.~G.\ 2003, \aj, 125, 2975




\bibitem{Lee2006} Lee, H., Skillman, E.~D., 
Cannon, J.~M., et al.\ 2006, \apj, 647, 970 

\bibitem{Lequeux1979} Lequeux, J., Peimbert, M., Rayo, J.~F., Serrano, A., \& Torres-Peimbert, S.\ 1979, \aap, 80, 155
%

\bibitem{McCall1985} McCall, M.~L., Rybski, 
P.~M., \& Shields, G.~A.\ 1985, \apjs, 57, 1 

\bibitem{Oppenheimer2010} Oppenheimer, B.~D., 
Dav{\'e}, R., Kere{\v s}, D., et al.\ 2010, \mnras, 406, 2325 


\bibitem{PMC2009} P{\'e}rez-Montero, E., \& Contini, T.\ 2009, \mnras, 398, 949

\bibitem{Petr2011} Petropoulou, V., 
V{\'{\i}}lchez, J., Iglesias-P{\'a}ramo, J., Papaderos, P., Magrini, L., 
Cedr{\'e}s, B., \& Reverte, D.\ 2011, \apj, 734, 32 



 \bibitem{Pettini2004} Pettini, M., \& Pagel, B.~E.~J.\ 2004, \mnras, 348, L59

\bibitem{Solanes2001} Solanes, J.~M., 
Manrique, A., Garc{\'{\i}}a-G{\'o}mez, C., et al.\ 2001, \apj, 548, 97 

\bibitem{Tecce2011} Tecce, T.~E., Cora, 
S.~A., \& Tissera, P.~B.\ 2011, \mnras, 416, 3170 

\bibitem{Tremonti2004} Tremonti, C.~A., et
 al.\ 2004, \apj, 613, 898


\bibitem{Vaduvescu2007} Vaduvescu, O.,
 McCall, M.~L., \& Richer, M.~G.\ 2007, \aj, 134, 604

\bibitem{Vaduvescu2011} Vaduvescu, O., Kehrig, C., Vilchez, J.~M., \& Unda-Sanzana, E.\ 2011, \aap, 533, A65 


\bibitem{vZH2006} van Zee, L., \& Haynes, M.~P.\ 2006, \apj, 636, 214

 \bibitem{Vilchez1995} Vilchez, J.~M.\ 1995, \aj,
 110, 1090

%
%
\end{thebibliography}
\end{document}